\documentclass[aps,prc,reprint,superscriptaddress,nofootinbib,longbibliography,floatfix]{revtex4-2}

\usepackage{graphicx}
\usepackage{dcolumn}
\usepackage{bm}
\usepackage{slashed}
\usepackage{amsfonts}
\usepackage{mathrsfs}
\usepackage{amssymb}
\usepackage{multirow}
\usepackage{subfigure}
\usepackage[T1]{fontenc}
\usepackage[english]{babel}
\usepackage{amsmath}
\usepackage{listings}
\usepackage{booktabs}
\usepackage{xcolor}
\usepackage{hyperref}

\begin{document}

\title{New perspective on cold fusion reactions: A microscopic description}

\author{Yinu Zhang}
\affiliation{Sino-French Institute of Nuclear Engineering and Technology, Sun Yat-sen University, Zhuhai 519082, China}

\author{Bo Han}
\affiliation{Sino-French Institute of Nuclear Engineering and Technology, Sun Yat-sen University, Zhuhai 519082, China}

\author{Yueping Fang}
\affiliation{Sino-French Institute of Nuclear Engineering and Technology, Sun Yat-sen University, Zhuhai 519082, China}

\author{Long Zhu}
\email{zhulong@mail.sysu.edu.cn}
\affiliation{Sino-French Institute of Nuclear Engineering and Technology, Sun Yat-sen University, Zhuhai 519082, China}

\begin{abstract}
A microscopic framework that combines the Hartree-Fock-Bogoliubov (HFB) approach with the fusion by diffusion (FBD) model is proposed to investigate the synthesis mechanism of superheavy nuclei (SHN). For the reaction $^{48}\text{Ca}+^{208}\text{Pb}$, the calculated evaporation-residue cross section (ERCS) reproduces the experimental data reasonably well. The method enables self-consistent extraction of the fusion injection point and inner barrier from HFB potential-energy surfaces (PES), thereby incorporating nuclear structure effects while eliminating phenomenological tuning at the fusion stage. For cold-fusion reactions, the PES features a hyperasymmetric valley driven by shell effects. This $^{208}$Pb anchored valley connects the entrance channel to compound nucleus formation and provides an exit channel for cluster decay. We further investigate the cold-fusion reactions $^{54}\text{Cr}+^{208}\text{Pb}$ and $^{58}\text{Fe}+^{208}\text{Pb}$, obtaining a near-exponential decrease of $P_{\text{CN}}$ with compound-nucleus charge $Z$, consistent with established systematics. This approach demonstrates a self-consistent framework that can reduce uncertainties in the fusion stage of SHN production.
\end{abstract}

\maketitle

\paragraph*{Introduction.}
The quest to synthesize superheavy nuclei (SHN) represents a frontier in modern nuclear physics. The motivations are not only the theoretical prediction of an island of stability, a region in the chart of nuclides where superheavy elements are expected to exhibit significantly longer half lives due to the stabilizing effects of closed nuclear shells, but also to answer one fundamental question: how many elements may exist~\cite{smitsQuestSuperheavyElements2023,giuliani2019colloquium,back2014recent, Hofmann2000}. 

Over the decades, significant progress has been achieved, and SHN from $Z=$110–118 were discovered~\cite{Hofmann2000,oganessian2007heaviest,morita2012new,oganessianSynthesisNewElement2010}. A primary focus in this field is the synthesis of new SHN. However, to synthesize the SHN beyond Og, the possible cross sections are lower than 1 pb. Consequently, theoretical predictions for optimal reaction systems and incident energies have become key factors~\cite{zhuLawOptimalIncident2023,gatesDiscoveryNewElements2024}.

Due to the complex dynamics of quantum many-body systems as well as the interplay with dissipation, the fusion process is the most unclear stage, which brings large uncertainties in predictions for synthesizing elements with $Z=119$ and 120~\cite{fang2024}.
A variety of theoretical models have been developed to describe fusion, including the dinuclear system model based on the master equation~\cite{XJBao2015,ZhuL2021} and the fusion-by-diffusion (FBD) model based on the Smoluchowski diffusion equation~\cite{swikatecki2005,cap2011,hagino2018,cap2022prcl,cap2022epja}. Although successful in many cases, these models still depend on phenomenological inputs, especially a potential-energy surface (PES) usually taken from a liquid drop model with microscopic corrections~\cite{moller2001}. Recent efforts have incorporated time-dependent Hartree-Fock (TDHF)~\cite{sekizawa2019,guolu2022,guolu2023} as a microscopic description of entrance channel capture dynamics and the FBD injection point. In Ref.~\cite{sekizawa2019}, a TDHF+Langevin scheme for hot fusion used the TDHF closest approach distance to initialize the diffusion stage; Refs.~\cite{guolu2022,guolu2023} applied the same strategy to cold- and hot-fusion systems and improved the microscopic grounding of $P_{\text{CN}}$. However, the inner-barrier height in these works was still taken from macroscopic--microscopic calculations.

The cold fusion reactions utilizing the compact target nuclei $^{208}\text{Pb}$ or $^{209}\text{Bi}$ play a significant role in synthesizing SHN. This approach is advantageous primarily due to the relatively low excitation energy of the formed compound nucleus, a direct consequence of the shell closure effects at $Z=82$ and $N=126$ in the target. Interestingly, the cluster decay in actinide and superheavy nuclei also reveals the strong influence of the double magic structure of $^{208}\text{Pb}$ as the daughter nucleus~\cite{poenaru2002}. The systematic investigations of cluster decay show that shell stabilized pathways lead nuclei to emit a cluster, heavier than an $\alpha$ particle but lighter than a typical fission fragment (A$\sim$100) and a daughter nucleus in the neighborhood of $^{208}\text{Pb}$. In superheavy nuclei, cluster decay is expected to be an important decay channel~\cite{poenaru2011,poenaru2012,zhanghongfei2014,warda2018,matheson2019}. The same shell stabilized, hyper-asymmetric valley on the microscopic PES manifests two distinct dynamical processes: cold fusion, which leads to the formation of a compound nucleus, and cluster decay, which leads to its disintegration. This valley is entered through the incoming "neck-zip" mechanism during the cold fusion entrance channel~\cite{cap2011}, and exited through the outgoing path of cluster decay toward scission~\cite{armbruster1997}. This supports the view that cold fusion is the inverse process of cluster decay. In this work, we refer to this double magic $^{208}\text{Pb}$ anchored valley as the cluster decay valley.

To self-consistently incorporate nuclear structure effects in the fusion dynamics, we introduce a novel hybrid framework that integrates the Hartree-Fock-Bogoliubov (HFB) approach and FBD, allowing the direct extraction of the injection point and inner fusion barrier from the microscopically calculated PES. The HFB framework provides such an approach by treating the mean field and pairing field on equal footing through the Bogoliubov transformation, simultaneously capturing single-particle dynamics and superfluidity~\cite{ringandschuck,bender2003}. Microscopic multidimensional PES calculations have been widely used in studies of fission~\cite{zdeb2021}, quasifission~\cite{simenel2021}, cluster decay~\cite{warda2011,warda2018,matheson2019,giuliani2023}. By incorporating shell and pairing effects that stabilize superheavy systems, they provide critical insights into both nuclear structure and fusion dynamics.


\paragraph*{Theoretical framework.}
The synthesis of SHN is commonly described as a sequence of three largely independent stages: (i) capture and formation of a dinuclear system, (ii) the fusion into a compact compound nucleus, and (iii) the de-excitation of the compound nucleus through particle evaporation to form a final SHN. 
Accordingly, the evaporation-residue cross section (ERCS) can be expressed as:
 \begin{equation}
     \sigma_{\text{ER}} \;=\; \sigma_{\text{capture}}\;\times\; P_{\text{CN}}\;\times\; W_{\text{surv}}
 \end{equation}



The capture cross section $\sigma_{\text{capture}}$ can be reasonably calculated using the Hill-Wheeler formula in combination with the empirical barrier distribution~\cite{wangSystematicsCaptureFusion2017}. The survival probability of the excited compound nucleus $W_{\text{surv}}$ in the process of its cooling by means of neutron evaporation and $\gamma$-emission in the competition with fission and emission of light charged particles is usually obtained from the statistical model, the detailed descriptions of which can be seen in Ref. ~\cite{Li2018}.


In contrast to the capture and survival stages, the fusion process is the primary source of uncertainty in ERCS calculations~\cite{shencaiwan2016}. Achieving an accurate description of $P_{\text{CN}}$ is therefore pivotal for reliable ERCS predictions. To calculate $P_{\text{CN}}$, the dynamical evolution can be described as follows. Immediately after the two nuclei touch, surface nucleons rapidly intermix on a very short timescale under the strong attractive interaction. A neck then forms quickly at an approximately fixed mass asymmetry and the overall elongation. In the FBD model, the ``neck-zip'' mechanism refers to the rapid growth of this neck degree of freedom, which progressively links the two fragments into a more compact composite configuration while largely preserving the entrance-channel mass asymmetry~\cite{cap2011}. During this fast relaxation stage, the system follows the minimum-energy valley on the PES, conveniently characterized by its projection onto the octupole moment $Q_{30}$. The endpoint of this neck-driven relaxation is identified as the injection configuration $S_{\text{inj}}$; from there, the system starts its thermally assisted uphill diffusion over the saddle point $S_{\text{sad}}$.
In order to fuse, the system must thermally diffuse uphill over an inner barrier of height $B$ with diffusion probability $P_{\text{diff}}$ to form a compact compound nucleus \cite{swikatecki2005,boilley2011,cap2022epja}. In this framework, the fusion probability is taken to be the diffusion probability, $P_{\text{CN}}\thickapprox P_{\text{diff}}$. 
In this way, $P_{\text{CN}}$ is given
\begin{equation}
P_{\text{CN}} = \frac{1}{2}(1-\mathrm{erf}(\sqrt{\frac{B}{T}} )),
\end{equation}
where $B=V(S_{\text{sad}})-V(S_{\text{inj}})$ is the height of the inner fusion barrier, and $T=\sqrt{E^{*}/a}$ is the temperature of the fusing system. Here, $a=A/12$ $\mathrm{MeV}^{-1}$. For synthesizing SHN, $P_{\text{CN}} \ll 1$, making it the dominant factor in the severe suppression of  cross sections, reducing the overall probability of success by several orders of magnitude. 
In this work, we employ constrained Skyrme HFB calculations to construct the PES of the nascent composite system. The resulting PES yields the fusion injection point and the height of the inner fusion barrier in a fully self-consistent way, which serves as input to the FBD for computing $P_{\text{CN}}$.



We integrate the HFB framework with the FBD in a self-consistent way. After crossing the Coulomb barrier, the projectile and target come into contact and reach the injection configuration, where a neck forms and rapidly develops between them. This process drives the system into the $^{208}$Pb shell effects induced asymmetric valley on the PES. A key application of the HFB is to provide the multi-dimensional PES for the compound system,  which dictates the energetically favorable pathways for the fusion process. The PES maps the total energy of the system under a set of constraints on its shape, using the constrained HFB framework based on the energy density functional (EDF), namely
\begin{equation}
\hat{\mathcal{H}}' = \hat{\mathcal{H}}
 - \sum_{\tau=N,Z} \lambda_{\tau}\,\hat{N}_{\tau}
 - \sum_{\mu=2,3,4} \lambda_{\mu}\,\hat{Q}_{\mu 0},
\end{equation}
where $\hat{\mathcal{H}}$ is the microscopic nuclear Hamiltonian; $\hat{Q}_{\mu0}$ is the multipole operator; $\lambda_{\tau}$ for particle number constraints, and $\lambda_{\mu}$ for multipole constraints $Q_{\mu0}$ . For the fusion process, the most relevant multipole operators include:\\
\textbf{Elongation: }The quadrupole moment $Q_{20}$ represents the distance between the centers of mass of the nascent fragments.\\
\textbf{Mass asymmetry:} The octupole moment $Q_{30}$ defines the mass ratio of the two fragments. \\
\textbf{Neck formation: }The hexadecapole $Q_{40}$ describes the development of the neck between the fragments.\\
By solving the HFB equation, the self-consistent calculation converges on a specific deformed shape, and the resulting total energy defines a point on the PES. This methodology bridges static nuclear structure models and the dynamic processes of fusion and cluster decay. 

The self-consistent PES calculation enables identification of the asymmetric fusion-cluster decay valley, which is a minimum energy trajectory that connects the separation projectile–target configuration to the post-contact configuration, passes through the saddle region, and finally merges into the compound nucleus. The essential FBD inputs, the injection point and the inner fusion barrier height $B$, emerge as intrinsic properties of the microscopic HFB energy landscape, rather than as arbitrarily fitted parameters. These properties correspond to the diffusing system's initial state after contact.  By integrating the PES with diffusion dynamics, we establish a fully self-consistent microscopic framework to calculate fusion reactions.

The SkM* EDF~\cite{skm1982} is employed to accurately reproduce fission barriers and other properties of deformed nuclei, making it a well-suited tool for studying nuclear fission and fusion processes in the present work. Pairing was treated using the Lipkin-Nogami approximation to avoid pairing collapse~\cite{Stoitsov2003}. The constrained HFB equations are solved using the axially symmetric code HFBTHO~\cite{hfbtho4}, which employs quadrupole, octupole, and hexadecapole deformations as collective degrees of freedom.


\paragraph*{Results and discussion.}
The reaction $^{48}\text{Ca}$+$^{208}\text{Pb}$ is chosen as the primary case study for this framework due to its unique and advantageous properties. Both the projectile $^{48}\text{Ca}$ and the target $^{208}\text{Pb}$ are doubly magic nuclei. The use of two double magic nuclei as reaction partners highlights the importance of the shell effect in the fusion process, which is the critical factor for the existence of the SHN.  
\begin{figure*}[h!]
\includegraphics[width=\textwidth]{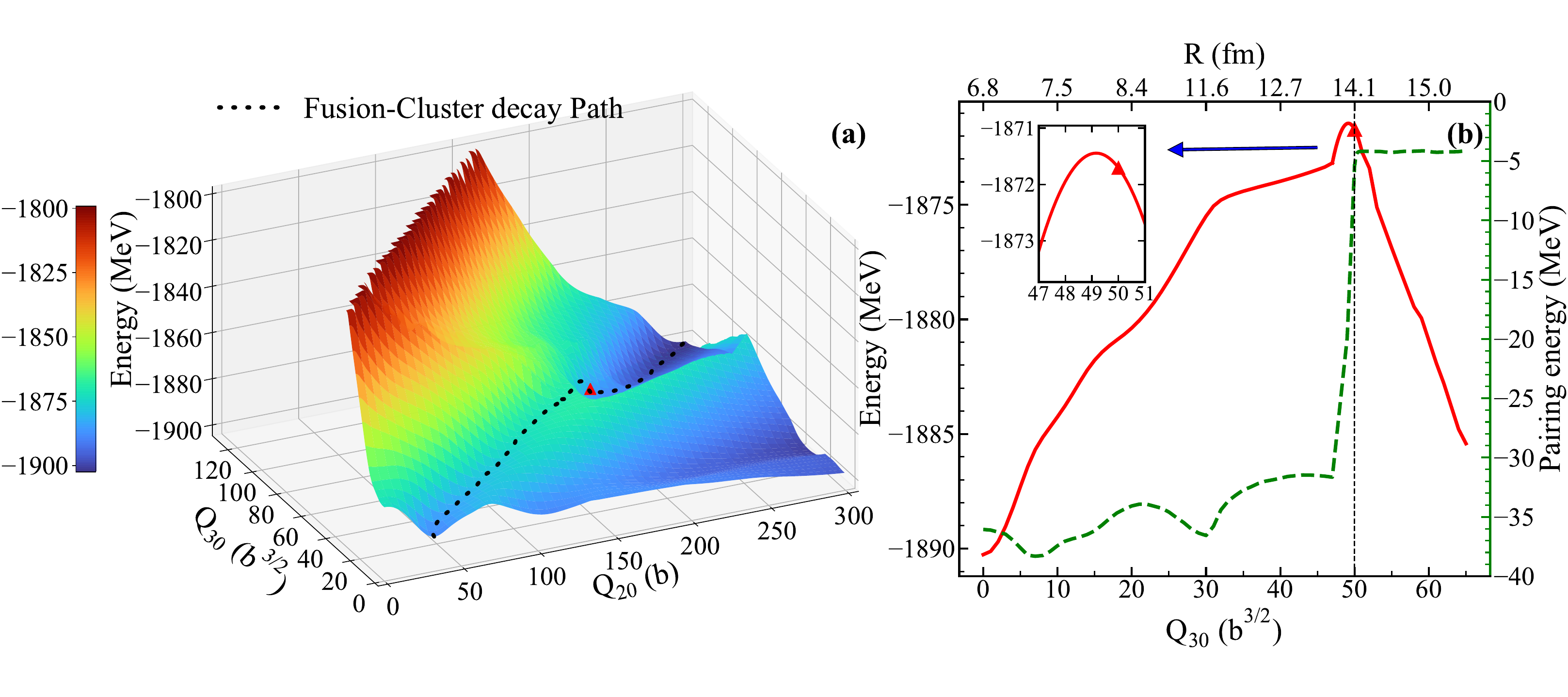}
\caption{\label{fig: pes} Landscape of the PES (a) of $^{256}\text{No}$ as a function of quadrupole moment $Q_{\text{20}}$ and octupole moment $Q_{\text{30}}$. (b) Hyper-asymmetric fusion-cluster decay path as functions of $Q_{\text{30}}$ (bottom) and the corresponding relative distance $R$ (top) between projectile and target nuclei. The red solid line denotes the total energy, the green dashed line denotes the pairing energy, and the red triangle marks the ``injection point''.
}
\end{figure*}

In Fig.~\ref{fig: pes}(a), the PES for the compound nucleus $^{256}\text{No}$ in the ($Q_{20}$, $Q_{30}$) plane is shown. A highly asymmetric fusion-cluster decay valley that connects the initial separated configuration to the fused compound nucleus can be observed. 
The PES is obtained from self-consistent HFB with constraints on the $Q_{20}$ for elongation, the $Q_{30}$ for mass asymmetry, and $Q_{40}$ for neck thickness. These coordinates provide a comprehensive description of the nuclear shape evolution during the fusion process. The injection point $\color{red}{\blacktriangle}$ (155 b, 50 b$^{3/2}$), which defines the initial configuration of the colliding projectile and target nuclei at the onset of the fusion stage, is self-consistently defined as the intersection point of the fusion path and the cluster decay path. The fusion path is traced from large separations by consistently considering the reaction involving $^{208}\text{Pb}$ and $^{48}\text{Ca}$ towards closer configurations, ensuring that the path aligns with the energetically favorable valley in the PES. Along this path, the potential energy increases as $Q_{20}$ and $Q_{30}$ decrease, reflecting the rising Coulomb repulsion and nuclear restructuring as the fragments approach the intersection point. 
The cluster decay path is traced along the valley of minimal energy from the ground state to the intersection point. In contrast to the traditional fission path, which is governed by $Q_{20}$, this path is driven by $Q_{30}$ as the primary collective coordinate due to the strong shell effects of $^{208}$Pb and mass asymmetry in the PES along this path. Such hyper-asymmetric cluster decay paths have been found in microscopic studies of cluster decays~\cite{warda2011,warda2018,giuliani2023,zhao2023,zhangyinu2026} by different EDFs. This criterion indicates the initial formation of a substantial nuclear bridge between the projectile and target, transitioning the system from a dinuclear to a more compact state, and eliminates the need for empirical parametrization of the injection distance. 
The inner fusion barrier height $B$ is then calculated as the energy difference between the saddle point and the injection point. Beyond the saddle point, the compound nucleus evolves toward its ground-state configuration along the minimum energy path, which corresponds to the reverse of the cluster decay path. This involves a further decrease in $Q_{20}$ and $Q_{30}$, as the system compacts and symmetrizes, minimizing the potential energy while incorporating shell and pairing effects self-consistently within the HFB framework.

In Fig.~\ref{fig: pes}(b), the asymmetric fusion-cluster decay path is projected onto $Q_{30}$ to provide an accurate description of the barrier height and shape for subsequent fusion calculations.
At the beginning, both $^{48}\text{Ca}$ and $^{208}$Pb are doubly magic, so pairing correlations are strongly suppressed by shell closures and the pairing energy is small.  As the dinuclear system forms a neck, the pairing energy experiences an enhancement, arising from the increased overlap of single-particle wave functions and the formation and reorganization of pairing correlations in the nascent compound nucleus. Pairing correlations can modify both the inner fusion barrier height~\cite{magierski2017,scamps2019,tong2022} and the fission barrier height~\cite{zdeb2025,wangxb2024}. This change in pairing energy may therefore contribute to shell stabilization and to a higher $P_{\text{CN}}$ by partially counteracting dissipative hindrance and quasi-fission competition. 
This $Q_{30}$ driven asymmetry path therefore bridges the fusion entrance channel and the cluster decay channel, providing a unified description that is essential for modeling fusion cross sections within our framework.
\begin{figure*}[h!]
\includegraphics[width=\textwidth]{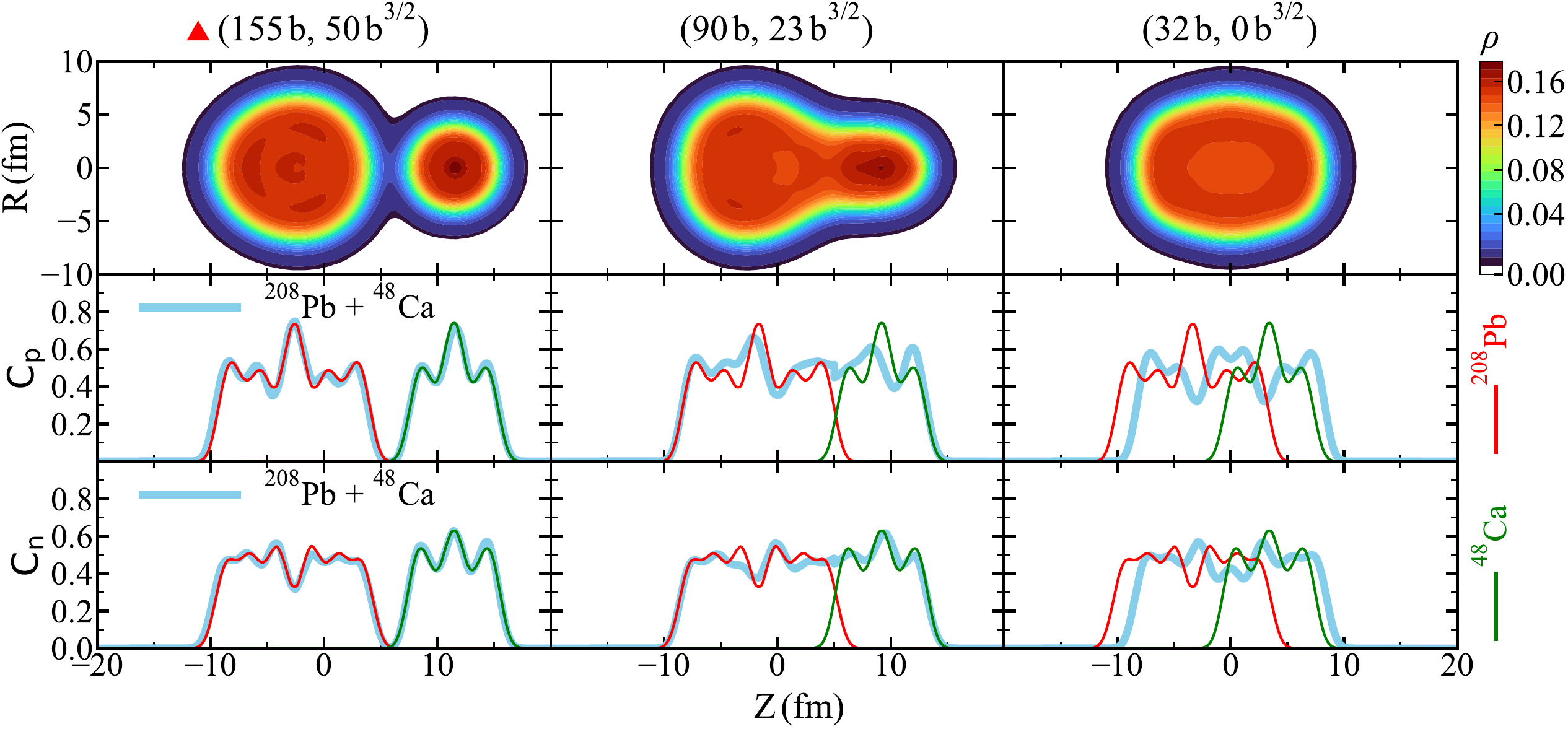}
\caption{\label{fig: density} Nucleonic densities (in nucleons/$\text{fm}^{3}$, top row) for $^{208}\text{Pb}+^{48}\text{Ca}$ obtained from HFB calculations with SkM* for three configurations along the hyper-asymmetric fusion-cluster decay path. The corresponding quadrupole $Q_{\text{20}}$ (b) and octupole moment $Q_{\text{30}}$ (b$^{3/2}$) are indicated in brackets. Proton (middle row) and neutron (bottom row) NLF profiles along the z axis (r = 0) for $^{256}\text{No}$ (blue thick line),  $^{208}\text{Pb}$ (red line) and $^{48}\text{Ca}$ (green line).}
\end{figure*}

To illustrate the microscopic dynamics of the $^{48}\text{Ca}$ + $^{208}\text{Pb}$ $\to$ $^{256}\text{No}$, we examine the evolution of the nuclear density distributions and the nucleon localization function (NLF) for proton and neutron along the asymmetric fusion-cluster decay pathway in Fig.~\ref{fig: density}. The NLF quantifies how localized nucleons are in space by measuring the likelihood of finding two nucleons with the same spin and isospin at a given position in a nucleus, providing a sensitive probe of clustering and shell structure in nuclei~\cite{reinhard2011}. Regions of enhanced localization correspond to the emergence of shell-like structures. A value of $\mathcal{C_{\tau}} \to 1$ indicates that nucleons occupy well-defined spatial regions, reflecting the quantum shell structure, and $\mathcal{C_{\tau}} \to 0.5$ corresponds to the homogeneous limit of the Fermi gas. The NLFs have been used to characterize the shell structures of nascent fragments in fissioning nuclei~\cite{scamps2018,zhangcl2016,sadhukhan2017,sadhukhan2020}, the cluster decay in SHN~\cite{matheson2019}, and the cluster structures in heavy-ion fusion reactions~\cite{schuetrumpf2017}. 
The first column corresponds to the injection point $\color{red}{\blacktriangle}$ with (155 b,  50 b$^{3/2}$) for $^{256}$No. As the projectile and target approach and initiate contact, which is marked by the formation of a nascent neck, the density distributions begin to elongate and overlap, leading to a gradual bridging in the neck region. The NLFs along the $z$-axis for the emerging composite system reveal a persistence of shell patterns akin to those of the $^{208}$Pb and $^{48}$Ca ground states, suggesting that initial shell effects from the reactants influence the early fusion stages and contribute to the stability of the dinuclear configuration. 
The second column shows that upon surpassing the inner fusion barrier and progressing toward fusion, the compound nucleus adopts a highly asymmetric shape, characterized by pronounced elongation and mass asymmetry. At this stage, the density distribution transitions into a pear-like shape, with the neck thickening and the system becoming more compact. The shell structures associated with $^{208}$Pb and $^{48}$Ca progressively fade, as indicated by the smoothing of the NLF peaks along the $z$-axis. This loss of NLF patterns reflects enhanced nucleon mixing and strengthened pairing correlations, which promote the formation of a unified $^{256}$No compound nucleus. 
The last column corresponds to the ground state configuration of $^{256}$No. While the density distributions exhibit no obvious internal substructures, the proton NLF displays  concentric elliptical ridges and the neutron NLF exhibits two pronounced maxima, consistent with the intrinsic prolate deformation. These visualizations underscore the role of shell evolution in SHN fusion, highlighting the microscopic fidelity of our framework. 

\begin{figure}[h!]
\includegraphics[width=\columnwidth]{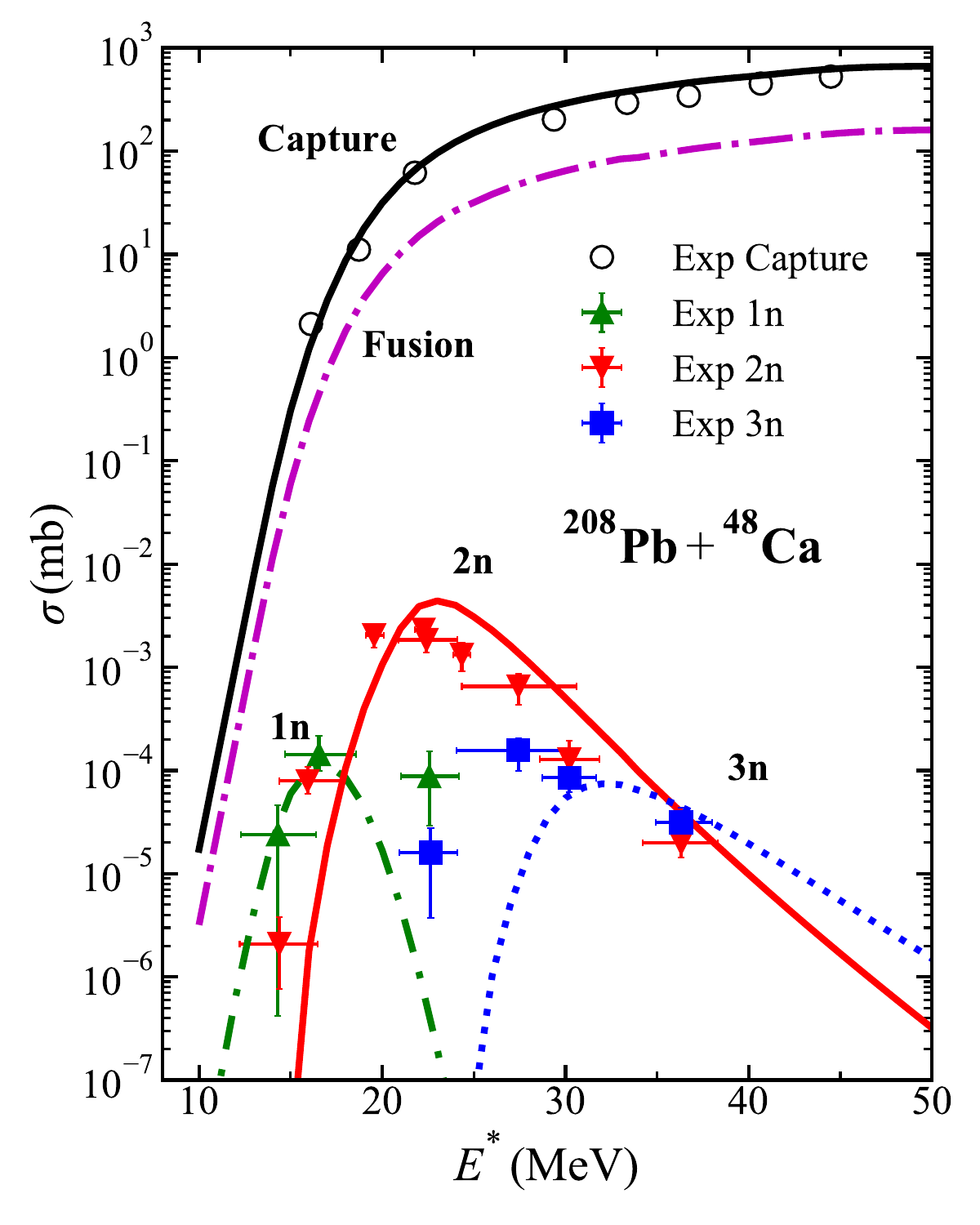}
\caption{\label{fig: crosssection} Energy dependence of the cross section for the synthesis of superheavy nuclei in cold fusion reactions. The excitation functions for the $1n$, $2n$, and $3n$ channels of the reaction $^{208}\text{Pb}(^{48}\text{Ca},xn)^{256-x}\text{No}$, calculated within the HFB+FBD framework in comparison with the experimental data~\cite{prokhorova2008fusion,gaggeler1989cold}.}
\end{figure}

To benchmark the HFB+FBD framework, the experimental capture and ERCS as a function of excitation energy in the reaction $^{208}\text{Pb}(^{48}\text{Ca},xn)^{256-x}\text{No}$ are compared with the calculations, as shown in Fig.~\ref{fig: crosssection}. The $Q$ value equals -152.87 MeV~\cite{Sierk1986,Huang2024}. The calculations can reproduce capture cross sections rather well with the Hill-Wheeler formula. 
The fusion cross sections are obtained by combining these capture cross sections with microscopically determined fusion probabilities $P_{\text{CN}}$, where both the injection configuration and the inner fusion barrier entering the diffusion dynamics are extracted self-consistently from the HFB PES. 
The capture and fusion cross sections remain comparable across different excitation energies, indicating a correspondingly high fusion probability. Within our microscopic treatment, the inner fusion barrier height is 
$B=$ 0.25 MeV. Since both $^{48}\text{Ca}$ and $^{208}\text{Pb}$ are doubly magic nuclei, their pronounced shell closures could enhance the stability of the system at the injection point (as demonstrated in Fig.~\ref{fig: density}), thereby lowering the inner fusion barrier height $B$ obtained from a microscopic self-consistent injection point. This reduction means an enhanced $P_{\text{CN}}$. We further employ the statistical model, which has been widely used in previous works, to calculate the survival probabilities of $^{256}$No in different evaporation channels. The calculated ERCS are in rather good agreement with the experimental data, thereby demonstrating the predictive effectiveness of the HFB+FBD framework.
\begin{figure}[h!]
\includegraphics[width=\columnwidth]{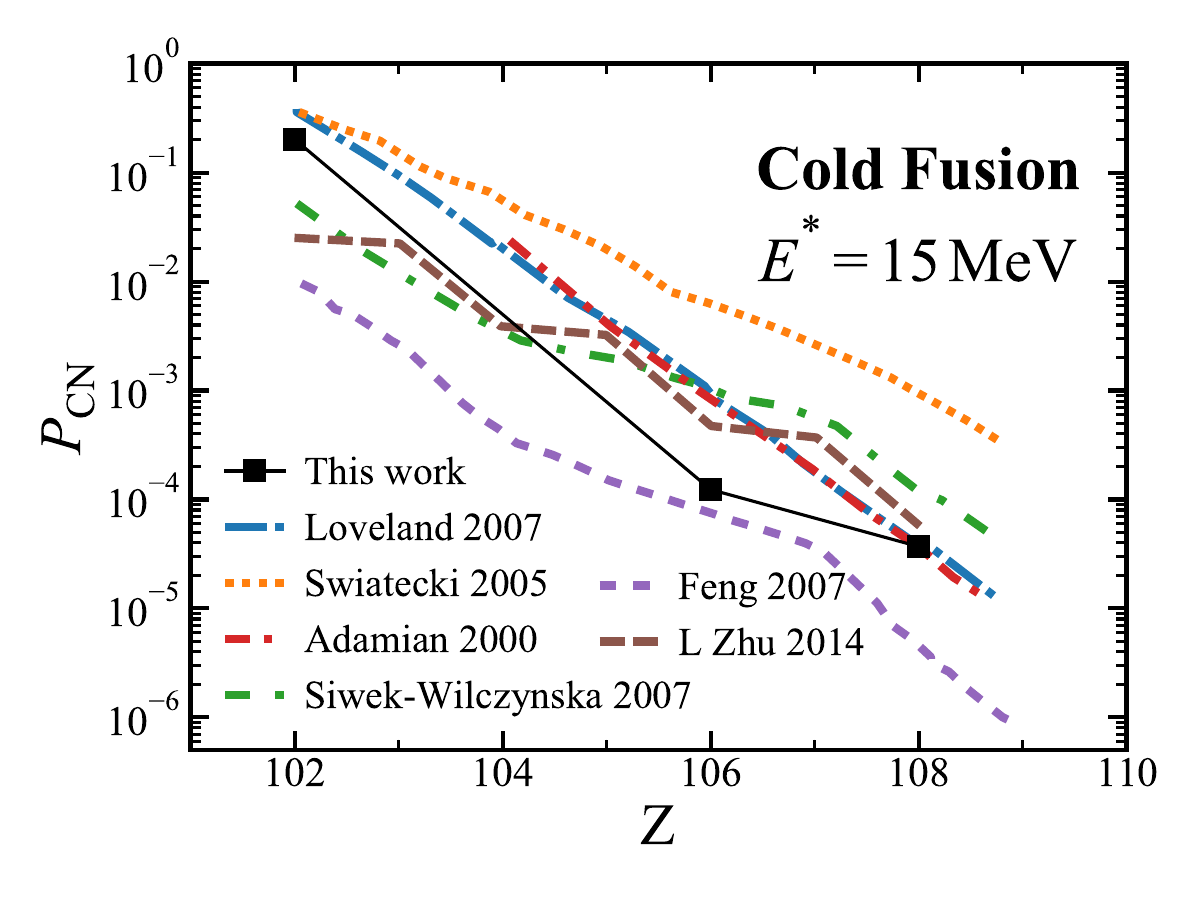}
\caption{\label{fig: Fig4_Pcn}Calculated fusion probability $P_{\text{CN}}$ for different cold fusion systems at $E^{{*}}=15\,{\text{MeV}}$, approximately the optimal energy of 1n channel. 
}
\end{figure}

To further validate the systematic applicability of the HFB+FBD framework, additional calculations were performed for the $^{54}\text{Cr}$ + $^{208}\text{Pb}$ and $^{58}\text{Fe}$ + $^{208}\text{Pb}$ reactions. As shown in Fig.~\ref{fig: Fig4_Pcn}, at an excitation energy of $E^{\text{*}}=15$ MeV, approximately the optimal energy for the 1n channel, the calculated fusion probability $P_{\text{CN}}$ decreases exponentially with increasing compound-nucleus charge number $Z$, revealing the characteristic fusion hindrance behavior of cold fusion reactions. A detailed comparison with the systematics reported by Loveland~\cite{lovelandSynthesisTransactinideNuclei2007}, Świątecki~\cite{swikatecki2005}, Siwek-Wilczyńska~\cite{siwek-wilczynskaROLEMASSASYMMETRY2007}, Adamian~\cite{adamianIsotopicDependenceFusion2000}, Feng~\cite{fengFormationSuperheavyNuclei2007}, and Zhu~\cite{zhuProductionCrossSections2014} shows that our predictions are reasonable, confirming that the proposed framework provides a consistent and physically sound description of the entrance channel dynamics governing compound-nucleus formation.

\paragraph*{Summary.}
We demonstrate a self-consistent framework that couples fully microscopic HFB nuclear structure calculations with FBD for fusion dynamics. By determining the fusion injection point and the inner fusion barrier height directly from the microscopically calculated PES, we enable a self-consistent calculation of the compound-nucleus formation probability $P_{\text{CN}}$. We choose $^{48}\text{Ca}$ + $^{208}\text{Pb}$ as the primary case study because both partners are doubly magic nuclei, a combination that maximizes the impact of shell effects, the critical stabilizing factor underlying the existence of SHN. The same shell stabilized, $Q_{30}$ dominated hyper-asymmetric valley on the microscopic PES underlies both cold fusion entry via ``neck‑zip'' mechanism and cluster decay exit toward scission, anchored by the doubly magic $^{208}\text{Pb}$. 
Moreover, pairing correlations play an important role in  $P_{\text{CN}}$. 
Using microscopic densities and the NLF, we track how shell structures evolve after contact and gradually dissipate along the fusion path. In future work, we will further investigate competition between pairing effects and finite-temperature effects. Our framework quantitatively reproduces ERCS in accordance with established systematics and predicts the observed near exponential decrease of $P_{\text{CN}}$ with $Z$ for the $^{48}\text{Ca}$, $^{54}\text{Cr}$, and $^{58}\text{Fe}$+$^{208}\text{Pb}$ reactions. This self-consistent approach reduces uncertainties in the fusion stage for cold fusion reactions. Its microscopic foundation facilitates the future incorporation of additional physics while preserving predictive reliability, thereby establishing a new way for theoretical descriptions of SHN synthesis.

\begin{acknowledgments}
This work was supported by the National Natural Science Foundation of China under Grants No. 12305129, No. 12475129, No. 12075327, and No. 12475136, and the Guangdong Major Project of Basic and Applied Basic Research under Grant No. 2021B0301030006.
\end{acknowledgments}




\bibliographystyle{apsrev4-2}
\bibliography{bibliography}

@article{warda2011,
  title = {Microscopic description of cluster radioactivity in actinide nuclei},
  author = {Warda, M. and Robledo, L. M.},
  journal = {Phys. Rev. C},
  volume = {84},
  issue = {4},
  pages = {044608},
  numpages = {17},
  year = {2011},
  month = {Oct},
  publisher = {American Physical Society},
  doi = {10.1103/PhysRevC.84.044608},
  url = {https://link.aps.org/doi/10.1103/PhysRevC.84.044608}
}

@article{zhao2023,
  title = {Microscopic description of $\ensuremath{\alpha}, 2\ensuremath{\alpha}$, and cluster decays of $^{216\ensuremath{-}220}\mathrm{Rn}$ and $^{220\ensuremath{-}224}\mathrm{Ra}$},
  author = {Zhao, J. and Ebran, J.-P. and Heitz, L. and Khan, E. and Mercier, F. and Nik\ifmmode \check{s}\else \v{s}\fi{}i\ifmmode \acute{c}\else \'{c}\fi{}, T. and Vretenar, D.},
  journal = {Phys. Rev. C},
  volume = {107},
  issue = {3},
  pages = {034311},
  numpages = {9},
  year = {2023},
  month = {Mar},
  publisher = {American Physical Society},
  doi = {10.1103/PhysRevC.107.034311},
  url = {https://link.aps.org/doi/10.1103/PhysRevC.107.034311}
}

@article{reinhard2011,
  title = {Localization in light nuclei},
  author = {Reinhard, P.-G. and Maruhn, J. A. and Umar, A. S. and Oberacker, V. E.},
  journal = {Phys. Rev. C},
  volume = {83},
  issue = {3},
  pages = {034312},
  numpages = {5},
  year = {2011},
  month = {Mar},
  publisher = {American Physical Society},
  doi = {10.1103/PhysRevC.83.034312},
  url = {https://link.aps.org/doi/10.1103/PhysRevC.83.034312}
}

@article{schuetrumpf2017,
  title = {Cluster formation in precompound nuclei in the time-dependent framework},
  author = {Schuetrumpf, B. and Nazarewicz, W.},
  journal = {Phys. Rev. C},
  volume = {96},
  issue = {6},
  pages = {064608},
  numpages = {8},
  year = {2017},
  month = {Dec},
  publisher = {American Physical Society},
  doi = {10.1103/PhysRevC.96.064608},
  url = {https://link.aps.org/doi/10.1103/PhysRevC.96.064608}
}

@article{scamps2018,
  title={Impact of pear-shaped fission fragments on mass-asymmetric fission in actinides},
  author={Scamps, Guillaume and Simenel, C{\'e}dric},
  journal={Nature},
  volume={564},
  number={7736},
  pages={382--385},
  year={2018},
  publisher={Nature Publishing Group UK London}
}

@article{zhangcl2016,
  title = {Nucleon localization and fragment formation in nuclear fission},
  author = {Zhang, C. L. and Schuetrumpf, B. and Nazarewicz, W.},
  journal = {Phys. Rev. C},
  volume = {94},
  issue = {6},
  pages = {064323},
  numpages = {6},
  year = {2016},
  month = {Dec},
  publisher = {American Physical Society},
  doi = {10.1103/PhysRevC.94.064323},
  url = {https://link.aps.org/doi/10.1103/PhysRevC.94.064323}
}

@article{hfbtho4,
title = {Axially-deformed solution of the {S}kyrme-{H}artree-{F}ock-{B}ogoliubov equations using the transformed harmonic oscillator basis (IV) hfbtho (v4.0): A new version of the program},
journal = {Computer Physics Communications},
volume = {276},
pages = {108367},
year = {2022},
issn = {0010-4655},
doi = {https://doi.org/10.1016/j.cpc.2022.108367},
url = {https://www.sciencedirect.com/science/article/pii/S0010465522000868},
author = {P. Marević and N. Schunck and E.M. Ney and R. {Navarro Pérez} and M. Verriere and J. O'Neal},
}

@article{skm1982,
title = {Towards a better parametrisation of Skyrme-like effective forces: A critical study of the {S}k{M} force},
journal = {Nuclear Physics A},
volume = {386},
number = {1},
pages = {79-100},
year = {1982},
issn = {0375-9474},
doi = {https://doi.org/10.1016/0375-9474(82)90403-1},
url = {https://www.sciencedirect.com/science/article/pii/0375947482904031},
author = {J. Bartel and P. Quentin and M. Brack and C. Guet and H.-B. Håkansson},
}

@article{Stoitsov2003,
  title = {Systematic study of deformed nuclei at the drip lines and beyond},
  author = {Stoitsov, M. V. and Dobaczewski, J. and Nazarewicz, W. and Pittel, S. and Dean, D. J.},
  journal = {Phys. Rev. C},
  volume = {68},
  issue = {5},
  pages = {054312},
  numpages = {11},
  year = {2003},
  month = {Nov},
  publisher = {American Physical Society},
  doi = {10.1103/PhysRevC.68.054312},
  url = {https://link.aps.org/doi/10.1103/PhysRevC.68.054312}
}

@article{smitsQuestSuperheavyElements2023,
  title={The quest for superheavy elements and the limit of the periodic table},
  author={Smits, Odile R and D{\"u}llmann, Christoph E and Indelicato, Paul and Nazarewicz, Witold and Schwerdtfeger, Peter},
  journal={Nature Reviews Physics},
  volume={6},
  number={2},
  pages={86--98},
  year={2024},
  publisher={Nature Publishing Group UK London}
}

@article{gatesDiscoveryNewElements2024,
  title = {Toward the {{Discovery}} of {{New Elements}}: {{Production}} of {{Livermorium}} ( {{Z}} = 116 ) with $^{50}\mathrm{Ti}$},
  shorttitle = {Toward the {{Discovery}} of {{New Elements}}},
  author = {Gates, J. M. and Orford, R. and Rudolph, D. and Appleton, C. and Barrios, B. M. and Benitez, J. Y. and Bordeau, M. and Botha, W. and Campbell, C. M. and Chadderton, J. and Chemey, A. T. and Clark, R. M. and Crawford, H. L. and Despotopulos, J. D. and Dorvaux, O. and Esker, N. E. and Fallon, P. and Folden, C. M. and Gall, B. J. P. and Garcia, F. H. and Golubev, P. and Gooding, J. A. and Grebo, M. and Gregorich, K. E. and Guerrero, M. and Henderson, R. A. and Herzberg, R.-D. and Hrabar, Y. and King, T. T. and Kireeff Covo, M. and Kirkland, A. S. and Kr{\"u}cken, R. and Leistenschneider, E. and Lykiardopoulou, E. M. and McCarthy, M. and Mildon, J. A. and {M{\"u}ller-Gatermann}, C. and Phair, L. and Pore, J. L. and Rice, E. and Rykaczewski, K. P. and Sammis, B. N. and Sarmiento, L. G. and Seweryniak, D. and Sharp, D. K. and Sinjari, A. and Steinegger, P. and Stoyer, M. A. and Szornel, J. M. and Thomas, K. and Todd, D. S. and Vo, P. and Watson, V. and Wooddy, P. T.},
  year = {2024},
  month = oct,
  journal = {Physical Review Letters},
  volume = {133},
  number = {17},
  pages = {172502}
}

@article{oganessianSynthesisNewElement2010,
  title = {Synthesis of a {{New Element}} with {{Atomic Number Z}} = 117},
  author = {Oganessian, {\relax Yu}. {\relax Ts}. and Abdullin, F. {\relax Sh}. and Bailey, P. D. and Benker, D. E. and Bennett, M. E. and Dmitriev, S. N. and Ezold, J. G. and Hamilton, J. H. and Henderson, R. A. and Itkis, M. G. and Lobanov, {\relax Yu}. V. and Mezentsev, A. N. and Moody, K. J. and Nelson, S. L. and Polyakov, A. N. and Porter, C. E. and Ramayya, A. V. and Riley, F. D. and Roberto, J. B. and Ryabinin, M. A. and Rykaczewski, K. P. and Sagaidak, R. N. and Shaughnessy, D. A. and Shirokovsky, I. V. and Stoyer, M. A. and Subbotin, V. G. and Sudowe, R. and Sukhov, A. M. and Tsyganov, {\relax Yu}. S. and Utyonkov, V. K. and Voinov, A. A. and Vostokin, G. K. and Wilk, P. A.},
  year = {2010},
  month = apr,
  journal = {Physical Review Letters},
  volume = {104},
  number = {14},
  pages = {142502}
}

@article{zhuLawOptimalIncident2023,
  title = {Law of Optimal Incident Energy for Synthesizing Superheavy Elements in Hot Fusion Reactions},
  author = {Zhu, Long},
  year = {2023},
  month = may,
  journal = {Physical Review Research},
  volume = {5},
  number = {2},
  pages = {L022030}
}

@article{fang2024,
  title = {Bayesian Uncertainty Quantification for Synthesizing Superheavy Elements},
  author = {Fang, Yueping and Gao, Zepeng and Zhang, Yinu and Liao, Zehong and Yang, Yu and Su, Jun and Zhu, Long},
  year = {2024},
  month = nov,
  journal = {Physics Letters B},
  volume = {858},
  pages = {139069}
}

@article{swikatecki2005,
  title={Fusion by diffusion. II. Synthesis of transfermium elements in cold fusion reactions},
  author={{\'S}wi{\k{a}}tecki, WJ and Siwek-Wilczy{\'n}ska, K and Wilczy{\'n}ski, J},
  journal={Physical Review C—Nuclear Physics},
  volume={71},
  number={1},
  pages={014602},
  year={2005},
  publisher={APS}
}

@article{cap2011,
  title = {Nucleus-nucleus cold fusion reactions analyzed with the $l$-dependent ``fusion by diffusion'' model},
  author = {Cap, T. and Siwek-Wilczy\ifmmode \acute{n}\else \'{n}\fi{}ska, K. and Wilczy\ifmmode \acute{n}\else \'{n}\fi{}ski, J.},
  journal = {Phys. Rev. C},
  volume = {83},
  issue = {5},
  pages = {054602},
  numpages = {10},
  year = {2011},
  month = {May},
  publisher = {American Physical Society},
}

@article{wangSystematicsCaptureFusion2017,
  title = {Systematics of Capture and Fusion Dynamics in Heavy-Ion Collisions},
  author = {Wang, Bing and Wen, Kai and Zhao, Wei-Juan and Zhao, En-Guang and Zhou, Shan-Gui},
  year = {2017},
  month = mar,
  journal = {Atomic Data and Nuclear Data Tables},
  volume = {114},
  pages = {281--370}
}

@article{shencaiwan2016,
  title = {Synthesis of superheavy elements: Uncertainty analysis to improve the predictive power of reaction models},
  author = {L\"u, Hongliang and Boilley, David and Abe, Yasuhisa and Shen, Caiwan},
  journal = {Phys. Rev. C},
  volume = {94},
  issue = {3},
  pages = {034616},
  numpages = {12},
  year = {2016},
  month = {Sep},
  publisher = {American Physical Society}
}

@article{boilley2011,
  title = {Fusion Hindrance of Heavy Ions: {{Role}} of the Neck},
  shorttitle = {Fusion Hindrance of Heavy Ions},
  author = {Boilley, David and L{\"u}, Hongliang and Shen, Caiwan and Abe, Yasuhisa and Giraud, Bertrand G.},
  year = {2011},
  month = nov,
  journal = {Physical Review C},
  volume = {84},
  number = {5},
  pages = {054608}
}

@article{lovelandSynthesisTransactinideNuclei2007,
  title = {Synthesis of Transactinide Nuclei Using Radioactive Beams},
  author = {Loveland, W.},
  year = {2007},
  month = jul,
  journal = {Physical Review C},
  volume = {76},
  number = {1},
  pages = {014612}
}

@article{siwek-wilczynskaROLEMASSASYMMETRY2007,
  title = {{{ROLE OF MASS ASYMMETRY IN FUSION OF SUPER-HEAVY NUCLEI}}},
  author = {{Siwek-Wilczy{\'n}ska}, K. and {Skwira-Chalot}, I. and Wilczy{\'n}ski, J.},
  year = {2007},
  month = feb,
  journal = {International Journal of Modern Physics E},
  volume = {16},
  number = {02},
  pages = {483--490}
}

@article{fengFormationSuperheavyNuclei2007,
  title = {Formation of Superheavy Nuclei in Cold Fusion Reactions},
  author = {Feng, Zhao-Qing and Jin, Gen-Ming and Li, Jun-Qing and Scheid, Werner},
  year = {2007},
  month = oct,
  journal = {Physical Review C},
  volume = {76},
  number = {4},
  pages = {044606}
}

@article{adamianIsotopicDependenceFusion2000,
  title = {Isotopic Dependence of Fusion Cross Sections in Reactions with Heavy Nuclei},
  author = {Adamian, G.G. and Antonenko, N.V. and Scheid, W.},
  year = {2000},
  month = sep,
  journal = {Nuclear Physics A},
  volume = {678},
  number = {1-2},
  pages = {24--38}
}

@article{zhuProductionCrossSections2014,
  title = {Production Cross Sections of Superheavy Elements {{Z}} = 119 and 120 in Hot Fusion Reactions},
  author = {Zhu, Long and Xie, Wen-Jie and Zhang, Feng-Shou},
  year = {2014},
  month = feb,
  journal = {Physical Review C},
  volume = {89},
  number = {2},
  pages = {024615}
}

@article{prokhorova2008fusion,
  title={The fusion-fission and quasi-fission processes in the reaction $^{48}\mathrm{Ca}$+$^{208}\mathrm{Pb}$ at energies near the coulomb barrier},
  author={Prokhorova, EV and Bogachev, AA and Itkis, MG and Itkis, IM and Knyazheva, GN and Kondratiev, NA and Kozulin, EM and Krupa, L and Oganessian, Yu Ts and Pokrovsky, IV and others},
  journal={Nuclear Physics A},
  volume={802},
  number={1-4},
  pages={45--66},
  year={2008},
  publisher={Elsevier}
}

@article{gaggeler1989cold,
  title={Cold fusion reactions with $^{48}\mathrm{Ca}$},
  author={G{\"a}ggeler, HW and Jost, DT and T{\"u}rler, A and Armbruster, P and Br{\"u}chle, W and Folger, H and He{\ss}berger, FP and Hofmann, S and M{\"u}nzenberg, G and Ninov, V and others},
  journal={Nuclear Physics A},
  volume={502},
  pages={561--570},
  year={1989},
  publisher={Elsevier}
}

@article{moller2001,
  title={Nuclear fission modes and fragment mass asymmetries in a five-dimensional deformation space},
  author={M{\"o}ller, Peter and Madland, DG and Sierk, AJ and Iwamoto, A},
  journal={Nature},
  volume={409},
  number={6822},
  pages={785--790},
  year={2001},
  publisher={Nature Publishing Group UK London}
}

@article{giuliani2019colloquium,
  title={Colloquium: Superheavy elements: Oganesson and beyond},
  author={Giuliani, Samuel A and Matheson, Zachary and Nazarewicz, Witold and Olsen, Erik and Reinhard, P-G and Sadhukhan, Jhilam and Schuetrumpf, Bastian and Schunck, Nicolas and Schwerdtfeger, Peter},
  journal={Reviews of Modern Physics},
  volume={91},
  number={1},
  pages={011001},
  year={2019},
  publisher={APS}
}

@article{back2014recent,
  title={Recent developments in heavy-ion fusion reactions},
  author={Back, BB and Esbensen, H and Jiang, CL and Rehm, KE},
  journal={Reviews of Modern Physics},
  volume={86},
  number={1},
  pages={317--360},
  year={2014},
  publisher={APS}
}

@article{Hofmann2000,
  title = {The discovery of the heaviest elements},
  author = {Hofmann, S. and M\"unzenberg, G.},
  journal = {Rev. Mod. Phys.},
  volume = {72},
  issue = {3},
  pages = {733--767},
  numpages = {0},
  year = {2000},
  month = {Jul},
  publisher = {American Physical Society}
}

@article{warda2018,
  title = {Cluster radioactivity in superheavy nuclei},
  author = {Warda, M. and Zdeb, A. and Robledo, L. M.},
  journal = {Phys. Rev. C},
  volume = {98},
  issue = {4},
  pages = {041602},
  numpages = {5},
  year = {2018},
  month = {Oct},
  publisher = {American Physical Society}
}

@article{Sierk1986,
  title = {Macroscopic model of rotating nuclei},
  author = {Sierk, Arnold J.},
  journal = {Phys. Rev. C},
  volume = {33},
  issue = {6},
  pages = {2039--2053},
  numpages = {0},
  year = {1986},
  month = {Jun},
  publisher = {American Physical Society},
  doi = {10.1103/PhysRevC.33.2039},
  url = {https://link.aps.org/doi/10.1103/PhysRevC.33.2039}
}

@article{Huang2024,
  title = {Multimodality of $^{187}\mathrm{Ir}$ fission studied by the Langevin approach},
  author = {Huang, Y. G. and Gu, F. C. and Feng, Y. J. and Wang, H. and Xiao, E. X. and Lei, X. and Zhu, L. and Su, J.},
  journal = {Phys. Rev. C},
  volume = {109},
  issue = {3},
  pages = {034609},
  numpages = {9},
  year = {2024},
  month = {Mar},
  publisher = {American Physical Society},
  doi = {10.1103/PhysRevC.109.034609},
  url = {https://link.aps.org/doi/10.1103/PhysRevC.109.034609}
}

@article{Li2018,
  title = {Predictions for the synthesis of superheavy elements {Z}=119 and 120},
  author = {Li, Fan and Zhu, Long and Wu, Zhi-Han and Yu, Xiao-Bin and Su, Jun and Guo, Chen-Chen},
  journal = {Phys. Rev. C},
  volume = {98},
  issue = {1},
  pages = {014618},
  numpages = {9},
  year = {2018},
  month = {Jul},
  publisher = {American Physical Society},
  doi = {10.1103/PhysRevC.98.014618},
  url = {https://link.aps.org/doi/10.1103/PhysRevC.98.014618}
}

@article{guolu2022,
  title = {Microscopic study of compound-nucleus formation in cold-fusion reactions},
  author = {Sun, Xiang-Xiang and Guo, Lu},
  journal = {Phys. Rev. C},
  volume = {105},
  issue = {5},
  pages = {054610},
  numpages = {9},
  year = {2022},
  month = {May},
  publisher = {American Physical Society},
  doi = {10.1103/PhysRevC.105.054610},
  url = {https://link.aps.org/doi/10.1103/PhysRevC.105.054610}
}

@article{guolu2023,
  title = {Microscopic study of the hot-fusion reaction $^{48}\mathrm{Ca}$+$^{238}\mathrm{U}$ with the constraints from time-dependent {Hartree}-{Fock} theory},
  author = {Sun, Xiang-Xiang and Guo, Lu},
  journal = {Phys. Rev. C},
  volume = {107},
  issue = {6},
  pages = {064609},
  numpages = {9},
  year = {2023},
  month = {Jun},
  publisher = {American Physical Society},
  doi = {10.1103/PhysRevC.107.064609},
  url = {https://link.aps.org/doi/10.1103/PhysRevC.107.064609}
}

@article{cap2022epja,
	title = {The {Fusion}-by-{Diffusion} model as a tool to calculate cross sections for the production of superheavy nuclei},
	volume = {58},
	issn = {1434-601X},
	url = {https://link.springer.com/10.1140/epja/s10050-022-00891-8},
	doi = {10.1140/epja/s10050-022-00891-8},
	language = {en},
	number = {11},
	urldate = {2025-08-17},
	journal = {The European Physical Journal A},
	author = {Cap, T. and Kowal, M. and Siwek-Wilczyńska, K.},
	month = nov,
	year = {2022},
	pages = {231},
}

@article{magierski2017,
  title = {Novel Role of Superfluidity in Low-Energy Nuclear Reactions},
  author = {Magierski, Piotr and Sekizawa, Kazuyuki and Wlaz\l{}owski, Gabriel},
  journal = {Phys. Rev. Lett.},
  volume = {119},
  issue = {4},
  pages = {042501},
  numpages = {6},
  year = {2017},
  month = {Jul},
  publisher = {American Physical Society},
  doi = {10.1103/PhysRevLett.119.042501},
  url = {https://link.aps.org/doi/10.1103/PhysRevLett.119.042501}
}

@article{tong2022,
  title = {Real space density-constrained time-dependent {H}artree-{F}ock-{B}ogoliubov method and pairing effects in the fusion of calcium isotopes},
  author = {Tong, Liang and Yan, Shiwei},
  journal = {Phys. Rev. C},
  volume = {105},
  issue = {1},
  pages = {014613},
  numpages = {9},
  year = {2022},
  month = {Jan},
  publisher = {American Physical Society},
  doi = {10.1103/PhysRevC.105.014613},
  url = {https://link.aps.org/doi/10.1103/PhysRevC.105.014613}
}

@article{sadhukhan2017,
  title = {Formation and distribution of fragments in the spontaneous fission of ${}^{\mathbf{240}}\mathbf{Pu}$},
  author = {Sadhukhan, Jhilam and Zhang, Chunli and Nazarewicz, Witold and Schunck, Nicolas},
  journal = {Phys. Rev. C},
  volume = {96},
  issue = {6},
  pages = {061301},
  numpages = {6},
  year = {2017},
  month = {Dec},
  publisher = {American Physical Society},
  doi = {10.1103/PhysRevC.96.061301},
  url = {https://link.aps.org/doi/10.1103/PhysRevC.96.061301}
}

@article{sadhukhan2020,
  title = {Efficient method for estimation of fission fragment yields of $r$-process nuclei},
  author = {Sadhukhan, Jhilam and Giuliani, Samuel A. and Matheson, Zachary and Nazarewicz, Witold},
  journal = {Phys. Rev. C},
  volume = {101},
  issue = {6},
  pages = {065803},
  numpages = {9},
  year = {2020},
  month = {Jun},
  publisher = {American Physical Society},
  doi = {10.1103/PhysRevC.101.065803},
  url = {https://link.aps.org/doi/10.1103/PhysRevC.101.065803}
}

@article{cap2022prcl,
  title = {Diffusion as a possible mechanism controlling the production of superheavy nuclei in cold fusion reactions},
  author = {Cap, T. and Kowal, M. and Siwek-Wilczy\ifmmode \acute{n}\else \'{n}\fi{}ska, K.},
  journal = {Phys. Rev. C},
  volume = {105},
  issue = {5},
  pages = {L051601},
  numpages = {6},
  year = {2022},
  month = {May},
  publisher = {American Physical Society},
  doi = {10.1103/PhysRevC.105.L051601},
  url = {https://link.aps.org/doi/10.1103/PhysRevC.105.L051601}
}

@article{bender2003,
  title = {Self-consistent mean-field models for nuclear structure},
  author = {Bender, Michael and Heenen, Paul-Henri and Reinhard, Paul-Gerhard},
  journal = {Rev. Mod. Phys.},
  volume = {75},
  issue = {1},
  pages = {121--180},
  numpages = {0},
  year = {2003},
  month = {Jan},
  publisher = {American Physical Society},
  doi = {10.1103/RevModPhys.75.121},
  url = {https://link.aps.org/doi/10.1103/RevModPhys.75.121}
}

@article{zdeb2021,
  title = {Description of the multidimensional potential-energy surface in fission of $^{252}\mathrm{Cf}$ and $^{258}\mathrm{No}$},
  author = {Zdeb, A. and Warda, M. and Robledo, L. M.},
  journal = {Phys. Rev. C},
  volume = {104},
  issue = {1},
  pages = {014610},
  numpages = {16},
  year = {2021},
  month = {Jul},
  publisher = {American Physical Society},
  doi = {10.1103/PhysRevC.104.014610},
  url = {https://link.aps.org/doi/10.1103/PhysRevC.104.014610}
}

@article{simenel2021,
  title={Comparison of fission and quasi-fission modes},
  author={Simenel, Cedric and McGlynn, Patrick and Umar, AS and Godbey, K},
  journal={Physics Letters B},
  volume={822},
  pages={136648},
  year={2021},
  publisher={Elsevier}
}

@book{ringandschuck,
  title={The nuclear many-body problem},
  author={Ring, Peter and Schuck, Peter},
  year={2004},
  publisher={Springer Science \& Business Media}
}

@article{matheson2019,
  title = {Cluster radioactivity of $_{118}^{294}\mathrm{Og}_{176}$},
  author = {Matheson, Zachary and Giuliani, Samuel A. and Nazarewicz, Witold and Sadhukhan, Jhilam and Schunck, Nicolas},
  journal = {Phys. Rev. C},
  volume = {99},
  issue = {4},
  pages = {041304},
  numpages = {6},
  year = {2019},
  month = {Apr},
  publisher = {American Physical Society},
  doi = {10.1103/PhysRevC.99.041304},
  url = {https://link.aps.org/doi/10.1103/PhysRevC.99.041304}
}

@article{giuliani2023,
  title={Cluster properties of heavy nuclei predicted with the Barcelona-Catania-Paris-Madrid energy density functional},
  author={Giuliani, Samuel A and Robledo, Luis M},
  journal={The European Physical Journal A},
  volume={59},
  number={12},
  pages={301},
  year={2023},
  publisher={Springer}
}

@article{armbruster1997,
  title={Heavy clusters in cold nuclear rearrangements in fusion and fission},
  author={Armbruster, P},
  journal={Il Nuovo Cimento A (1971-1996)},
  volume={110},
  number={9},
  pages={1111--1126},
  year={1997},
  publisher={Springer}
}

@article{hagino2018,
  title = {Hot fusion reactions with deformed nuclei for synthesis of superheavy nuclei: An extension of the fusion-by-diffusion model},
  author = {Hagino, K.},
  journal = {Phys. Rev. C},
  volume = {98},
  issue = {1},
  pages = {014607},
  numpages = {9},
  year = {2018},
  month = {Jul},
  publisher = {American Physical Society},
  doi = {10.1103/PhysRevC.98.014607},
  url = {https://link.aps.org/doi/10.1103/PhysRevC.98.014607}
}

@article{poenaru2002,
  title = {Systematics of cluster decay modes},
  author = {Poenaru, D. N. and Nagame, Y. and Gherghescu, R. A. and Greiner, W.},
  journal = {Phys. Rev. C},
  volume = {65},
  issue = {5},
  pages = {054308},
  numpages = {6},
  year = {2002},
  month = {Apr},
  publisher = {American Physical Society},
  doi = {10.1103/PhysRevC.65.054308},
  url = {https://link.aps.org/doi/10.1103/PhysRevC.65.054308}
}

@article{poenaru2011,
  title = {Heavy-Particle Radioactivity of Superheavy Nuclei},
  author = {Poenaru, D. N. and Gherghescu, R. A. and Greiner, W.},
  journal = {Phys. Rev. Lett.},
  volume = {107},
  issue = {6},
  pages = {062503},
  numpages = {4},
  year = {2011},
  month = {Aug},
  publisher = {American Physical Society},
  doi = {10.1103/PhysRevLett.107.062503},
  url = {https://link.aps.org/doi/10.1103/PhysRevLett.107.062503}
}

@article{poenaru2012,
  title = {Cluster decay of superheavy nuclei},
  author = {Poenaru, D. N. and Gherghescu, R. A. and Greiner, W.},
  journal = {Phys. Rev. C},
  volume = {85},
  issue = {3},
  pages = {034615},
  numpages = {7},
  year = {2012},
  month = {Mar},
  publisher = {American Physical Society},
  doi = {10.1103/PhysRevC.85.034615},
  url = {https://link.aps.org/doi/10.1103/PhysRevC.85.034615}
}

@article{zhanghongfei2014,
  title = {Competition between $\ensuremath{\alpha}$ decay and cluster radioactivity for superheavy nuclei with a universal decay-law formula},
  author = {Bao, Xiao Jun and Zhang, Hai Fei and Dong, Jian Min and Li, Jun Qing and Zhang, Hong Fei},
  journal = {Phys. Rev. C},
  volume = {89},
  issue = {6},
  pages = {067301},
  numpages = {5},
  year = {2014},
  month = {Jun},
  publisher = {American Physical Society},
  doi = {10.1103/PhysRevC.89.067301},
  url = {https://link.aps.org/doi/10.1103/PhysRevC.89.067301}
}

@article{oganessian2007heaviest,
  title={Heaviest nuclei from $^{48}\mathrm{Ca}$-induced reactions},
  author={Oganessian, Yuri},
  journal={Journal of Physics G: Nuclear and Particle Physics},
  volume={34},
  number={4},
  pages={R165},
  year={2007},
  publisher={IOP Publishing}
}

@article{morita2012new,
  title={New result in the production and decay of an isotope, $^{278}113$, of the 113th element},
  author={Morita, Kosuke and Morimoto, Kouji and Kaji, Daiya and Haba, Hiromitsu and Ozeki, Kazutaka and Kudou, Yuki and Sumita, Takayuki and Wakabayashi, Yasuo and Yoneda, Akira and Tanaka, Kengo and others},
  journal={Journal of the Physical Society of Japan},
  volume={81},
  number={10},
  pages={103201},
  year={2012},
  publisher={The Physical Society of Japan}
}

@article{ZhuL2021,
  title = {Unified description of fusion and multinucleon transfer processes within the dinuclear system model},
  author = {Zhu, Long and Su, Jun},
  journal = {Phys. Rev. C},
  volume = {104},
  issue = {4},
  pages = {044606},
  numpages = {8},
  year = {2021},
  month = {Oct},
  publisher = {American Physical Society},
}

@article{XJBao2015,
  title = {Theoretical study of the synthesis of superheavy nuclei using radioactive beams},
  author = {Bao, X. J. and Gao, Y. and Li, J. Q. and Zhang, H. F.},
  journal = {Phys. Rev. C},
  volume = {91},
  issue = {6},
  pages = {064612},
  numpages = {4},
  year = {2015},
  month = {Jun},
  publisher = {American Physical Society},
  doi = {10.1103/PhysRevC.91.064612},
  url = {https://link.aps.org/doi/10.1103/PhysRevC.91.064612}
}

@article{zhangyinu2026,
title = {Cluster-liquid phase separation in cluster decay},
journal = {Physics Letters B},
volume = {873},
pages = {140155},
year = {2026},
issn = {0370-2693},
doi = {https://doi.org/10.1016/j.physletb.2026.140155},
url = {https://www.sciencedirect.com/science/article/pii/S0370269326000092},
author = {Yinu Zhang and Long Zhu and Cenxi Yuan and Zepeng Gao}
}

@article{sekizawa2019,
  title = {Time-dependent Hartree-Fock plus Langevin approach for hot fusion reactions to synthesize the $Z=120$ superheavy element},
  author = {Sekizawa, K. and Hagino, K.},
  journal = {Phys. Rev. C},
  volume = {99},
  issue = {5},
  pages = {051602},
  numpages = {5},
  year = {2019},
  month = {May},
  publisher = {American Physical Society},
  doi = {10.1103/PhysRevC.99.051602},
  url = {https://link.aps.org/doi/10.1103/PhysRevC.99.051602}
}

@article{scamps2019,
  title = {Density-constrained time-dependent Hartree-Fock-Bogoliubov method},
  author = {Scamps, Guillaume and Hashimoto, Yukio},
  journal = {Phys. Rev. C},
  volume = {100},
  issue = {2},
  pages = {024623},
  numpages = {8},
  year = {2019},
  month = {Aug},
  publisher = {American Physical Society},
  doi = {10.1103/PhysRevC.100.024623},
  url = {https://link.aps.org/doi/10.1103/PhysRevC.100.024623}
}

@article{zdeb2025,
  title = {Microscopic pairing in fission dynamics},
  author = {Zdeb, A. and Baran, A. and Giuliani, S. A. and Robledo, L. M. and Warda, M.},
  journal = {Phys. Rev. C},
  volume = {112},
  issue = {4},
  pages = {044602},
  numpages = {11},
  year = {2025},
  month = {Oct},
  publisher = {American Physical Society},
  doi = {10.1103/w2mk-7jcv},
  url = {https://link.aps.org/doi/10.1103/w2mk-7jcv}
}

@article{wangxb2024,
  title={Sensitivity impacts owing to the variations in the type of zero-range pairing forces on the fission properties using the density functional theory},
  author={Su, Yang and Li, Ze-Yu and Liu, Li-Le and Dong, Guo-Xiang and Wang, Xiao-Bao and Chen, Yong-Jing},
  journal={Nuclear Science and Techniques},
  volume={35},
  number={3},
  pages={62},
  year={2024},
  doi={10.1007/s41365-024-01422-4},
  url = {https://link.springer.com/article/10.1007/s41365-024-01422-4},
  publisher={Springer}
}






\end{document}